# Optimized multi-site local orbitals in the large-scale DFT program CONQUEST

*Ayako Nakata,*[*,a] *David R. Bowler,*[b,c,d] *and Tsuyoshi Miyazaki*[**,b,e]

[a] International Center for Young Scientists (ICYS), National Institute for Materials Science (NIMS), 1-1 Namiki, Tsukuba, Ibaraki 305-0044, Japan

[b] Computational Materials Science Unit (CMSU), National Institute for Materials Science (NIMS), 1-1 Namiki, Tsukuba, Ibaraki 305-0044, Japan

[c] Department of Physics & Astronomy, University College London, Gower St, London WC1E 6BT, UK

[d] WPI-MANA, National Institute for Materials Science (NIMS), 1-1 Namiki, Tsukuba, Ibaraki 305-0044, Japan

[e] London Centre for Nanotechnology, University College London, 17-19 Gordon Street, London WC1H 0AH, UK

ABSTRACT

We introduce numerical optimization of multi-site support functions in the linear-scaling DFT code CONQUEST. Multi-site support functions, which are linear combinations of pseudo-atomic orbitals on a target atom and those neighbours within a cutoff, have been recently proposed to reduce the number of support functions to the minimal basis while keeping the accuracy of a large basis [*J. Chem. Theory Comput.*, 2014, **10**, 4813]. The coefficients were determined by using the local filter diagonalization (LFD) method [*Phys. Rev. B*, 2009, **80**, 205104]. We analyse the effect of numerical optimization of the coefficients produced by the LFD method. Tests on crystalline silicon, a benzene molecule and hydrated DNA systems show that the optimization improves the accuracy of the multi-site support functions with small cutoffs. It is also confirmed that the optimization guarantees the variational energy minimizations with multi-site support functions.

1. **Introduction**

Recent progress in theory and computing power has enabled us to simulate properties of condensed phase materials and molecules precisely with condensed-matter physics and quantum chemistry techniques. Density functional theory (DFT) is one of the most widely used tools for these simulations, because of its relatively low computational cost and the reasonable accuracy achieved by taking into account the electron correlations through the exchange-correlation functional. However, most DFT calculations have been performed on systems which contain only up to about a thousand atoms, because the computational cost scales cubically with the number of atoms in the system, $N$.

Our own CONQUEST code[1–3] is a DFT code for large-scale systems with real-space local orbital basis functions (called "support functions" in CONQUEST). Since the support functions are localized in finite regions, the matrices in the support-function basis are sparse, and sparse matrix multiplications have high parallel efficiency in CONQUEST.[4] CONQUEST supports both exact diagonalization (O($N^3$)) and linear-scaling (O($N$)) approaches to optimize the electronic structure, and the use of local orbitals reduces the computational cost in both methods. Recently, CONQUEST has succeeded in performing calculations on the systems including more than a million atoms with the O($N$) method.[3,5]

The support functions in CONQUEST are constructed as linear combinations of given basis functions. Two kinds of basis functions, b-spline (blip) finite-element basis functions[6] akin to plane-waves, and pseudo atomic orbital (PAO) basis functions[7] are used. We focus on the use of PAOs in this study. PAOs are the atomic-orbital basis functions found from the pseudo-potentials and consist of the radial functions multiplied by spherical harmonic functions.[8,9] Radial functions are described by numerical functions on regular grids. Although it is difficult to improve the accuracy of PAOs systematically, the accuracy of calculations with PAOs is usually improved by increasing the number of radial functions for each spherical harmonic function. PAOs with several radial functions for each spherical harmonic function are called "multiple-$\zeta$" PAOs, while the PAOs in which only one radial function is used for each spherical function are called "single-$\zeta$" PAOs. Since the computational cost depends cubically on the number of support functions in both diagonalization and O($N$) calculations, linear combinations of multiple-$\zeta$ PAOs are often taken to contract support functions. The PAOs without contractions are called "primitive" support functions.

The accuracy of the "contracted" support functions depends on the linear-combination coefficients. The coefficients are fixed to some optimized values in conventional contracted basis set in quantum chemistry.[10] On the other hand, in the contracted support functions in CONQUEST, the coefficients are optimized for each atom in each target system.[7] A similar contraction method was proposed by Ozaki and Kino.[11,12] Since the linear combinations are taken only with PAOs which are centred at the target atom, the conventional "single-site" support functions have to keep the point-group symmetry of the target atom. This constraint leads a limitation in reducing the number of support functions.[7]

We have recently proposed the "multi-site" support functions, which are the linear combinations of the PAOs on both target atoms and their neighbouring atoms in finite regions.[13] As they correspond to local molecular orbitals (MOs), the multi-site support functions are free from the limitation from the atomic orbital symmetry and can be reduced to the minimal basis size. To determine the linear-combination coefficients, we have applied the localized filter diagonalization (LFD) method which was proposed by Rayson and Briddon.[14,15] In the LFD method, the linear-combination coefficients are determined efficiently by using the local MO coefficients projected onto localized trial vectors. However, the energy minimizations with this projection are not variational.[13] This lack of variational freedom only causes serious problems when there are not enough neighbouring atoms included in the multi-site support functions. However, it is important to remove this problem to guarantee stable and accurate geometry optimizations and molecular dynamics simulations. Another benefit of the optimization of the coefficients is that the neighbour region to construct the multi-site support functions with reasonable accuracy will be reduced. The reduction of the support function region is one of the critical factors to save computational cost in CONQUEST.

In the present study, we assess the dependence on the neighbour atoms and the variational behaviour of multi-site support functions. Based on the assessment, we introduce numerical optimizations to guarantee the variational principles and stable calculations with multi-site support functions. In the next section we explain the method of multi-site support functions and its optimizations. The performance of the optimized multi-site support functions are assessed by analysing energy-volume curves, atomic forces and density of states for crystalline silicon (Si), a benzene molecule and hydrated DNA systems in the third section. The stability of the calculations is also investigated. The final section gives the conclusion of the present study.

## 2. Theory and computational details

The Kohn-Sham density matrix $\rho$ in DFT is defined as

$$\rho(\mathbf{r},\mathbf{r}') = \sum_n f_n \psi_n(\mathbf{r}) \psi_n(\mathbf{r}')^*, \quad (1)$$

where $\psi_n$ and $f_n$ are the $n$th Kohn-Sham orbitals and its occupation numbers. In CONQUEST, $\rho$ is expressed by support functions $\phi_{i\alpha}$ as

$$\rho(\mathbf{r},\mathbf{r}') = \sum_{i\alpha,j\beta} \phi_{i\alpha}(\mathbf{r}) K_{i\alpha,j\beta} \phi_{j\beta}(\mathbf{r}')^*. \quad (2)$$

**K** is the density matrix in support function basis. $i, j$ and $\alpha, \beta$ are the indices of atoms and support functions, respectively. The conventional single-site support functions which consist only of PAOs on the target atom is

$$\phi_{i\alpha}(\mathbf{r}) = \sum_\mu c_{i\alpha,i\mu} \chi_{i\mu}(\mathbf{r}). \quad (3)$$

**c** is the linear-combination coefficient and $\chi_{i\mu}$ is $\mu$th PAO on atom $i$. The coefficients **c** have been optimized to minimize the electronic energy.[7] On the other hand, the multi-site support function[13] is defined as the linear combination of the PAOs on not only the target atom but also on the neighbour atoms,

$$\phi_{i\alpha}(\mathbf{r}) = \sum_k^{neighbors} \sum_{\mu \in k} C_{i\alpha,k\mu} \chi_{k\mu}(\mathbf{r}), \quad (4)$$

where $k$ runs over the neighbour atoms which are within the radius of the multi-site region, $r_{MS}$, from atom $i$. The coefficients **C** in Eq. (4) have been determined by the LFD method.[13–15] In this method, the subspaces of Hamiltonian $\mathbf{H}_S$ and overlap matrices $\mathbf{S}_S$, which belong to the target atom and its neighbour atoms in the local diagonalization region of the radius $r_{LD}$, are constructed from the original Hamiltonian and overlap matrices in PAO basis $\mathbf{H}^{PAO}$ and $\mathbf{S}^{PAO}$. The local diagonalization with $\mathbf{H}_S$ and $\mathbf{S}_S$ yields the eigenvectors $\mathbf{C}_S$ and eigenvalues $\varepsilon_S$.

$$\mathbf{H}_s \mathbf{C}_s = \varepsilon_s \mathbf{S}_s \mathbf{C}_s. \tag{5}$$

$\mathbf{C}_S$ is localized by the projection on the trial vectors $\mathbf{t}$ which are localized around the target atom.

$$\mathbf{C'} = \mathbf{C}_s f(\varepsilon_s) \mathbf{C}_s^T \mathbf{S}_s \mathbf{t}. \tag{6}$$

PAOs on the target atom are used as trial vectors in the present study. $f(\varepsilon)$ is the Fermi-Dirac function with the chemical potential close to the Fermi level, which eliminates the effect from unoccupied MOs in high energy regions. $\mathbf{C'}$ is mapped to the corresponding positions in $\mathbf{C}$ in Eq. (4).

We can use double cutoffs, i.e., use different values for $r_{MS}$ and $r_{LD}$. The use of larger $r_{LD}$ improves the accuracy of the contraction generally, while smaller $r_{MS}$ is desirable to save the computational cost. We have confirmed that the use of larger $r_{LD}$ with fixed $r_{MS}$ tends to improve the accuracy, especially for the description of unoccupied band structures.[13]

Once $\mathbf{C}$ is determined, a self-consistent-field (SCF) calculation is performed. This calculation is variational when $\mathbf{C}$ is unchanged during the SCF procedure. After the SCF calculation converges, we update $\mathbf{C}$ by using the converged Hamiltonian. This update changes the support function space. Therefore, the two-step procedure, the SCF calculations and the subsequent update of $\mathbf{C}$, is repeated until the energy and density converge. If the accuracy of the multi-site support functions is the same as that of the primitive support functions, it is guaranteed that there exists a set of the PAO coefficients $\mathbf{C}$ which provides the SCF charge density giving the Hamiltonian consistent with the PAO coefficients. However, if the accuracy of the multi-site support functions is not sufficient, for instance if **C** is calculated simply by the projection method as explained above, there is no guarantee that we can obtain consistent multi-site support functions and SCF charge density.

This inconsistency can be avoided if **C** is determined by the numerical optimization method such as the conjugate gradient method. We need the gradient of the electronic energy with respect to **C** to perform the numerical optimization. The gradient with respect to the coefficients can be calculated as the partial derivative through support functions as

$$\frac{\partial E}{\partial C_{i\alpha,k\mu}} = \frac{\partial E}{\partial \phi_{i\alpha}} \frac{\partial \phi_{i\alpha}}{\partial C_{i\alpha,k\mu}} = \frac{\partial E}{\partial \phi_{i\alpha}} \chi_{k\mu}. \tag{7}$$

The gradient with respect to $\phi_{i\alpha}$ is obtained as

$$\frac{\partial E}{\partial \phi_{i\alpha}(\mathbf{r})} = 4\sum_{\beta}\left[K_{\alpha\beta}\hat{H} + G_{\alpha\beta}\right]\phi_{j\beta}(\mathbf{r}), \tag{8}$$

where **G** is given as the energy-weighted density matrix

$$G_{\alpha\beta} = \sum_{n} f_n \varepsilon_n u_{n\alpha} u_{n\beta}^*, \tag{9}$$

in the diagonalization calculations and

$$G_{\alpha\beta} = 3(LHL)_{\alpha\beta} - 2(LSLHL + LHLSL)_{\alpha\beta}, \tag{10}$$

in the O(N) calculations. **L** is the auxiliary density matrix which has the relationship with **K** under the idempotency condition[16] as,

$$K = 3LSL - 2LSLSL. \qquad (11)$$

The precise derivation of Eqs. (8) – (10) is shown in reference [17]. The way to calculate the gradient with respect to the PAO coefficients in multi-site support functions is not significantly different from that in single-site support functions. The only point that we should note is that the atom index of PAOs $k$ in Eq. (7) runs over not only the target atom $i$ but also the neighbouring atoms of $i$.

It might be difficult to determine **C** using only numerical optimization, because the number of the PAO coefficients in multi-site support functions, which depends on the multi-site region $r_{MS}$, is usually much larger than that of single-site support functions. Therefore, we first obtain **C** by performing the two-step calculations with the LFD method, and use the **C** obtained as the initial values for the numerical optimization.

### 3. Results and discussions

*3.1. Energy-volume curves of crystalline silicon*

First, we have performed calculations on crystalline Si in order to assess the performance of the multi-site support functions with and without the optimization of their PAO coefficients. The purpose of this assessment is to find how precisely the multi-site support functions can reproduce the results by primitive support functions, which provide the best values in the given PAO space, and how the optimization of the coefficients affects the results.

Valence triple $\zeta$ plus double polarization (TZDP) PAOs[18] are generated using the Siesta code.[9] A multi-site support function on a Si atom consists of all of the PAOs on the neighbour Si atoms in the multi-site range $r_{MS}$. The numbers of primitive and multi-site support functions for each Si atom are 22 (=3$s$, 3$p$, 2$d$) and four, respectively. The multi-site support functions with and without the coefficient optimization after the LFD calculations are denoted as ($r_{LD}$-$r_{MS}$) and ($r_{LD}$-$r_{MS}$)opt hereafter, where $r_{LD}$ and $r_{MS}$ are in bohr. The local density approximation (LDA)[19] functional is used. We use exact diagonalization, not O($N$) method, in the present study to concentrate on the accuracy and efficiency of the multi-site support functions. The number of $k$-points used in the diagonalization method for bulk Si is (4, 4, 4) with a Monkhorst-Pack mesh.

Figure 1 shows the energy-volume curves of crystalline Si calculated with multi-site support functions with several $r_{MS}$. $r_{LD}$ is set to be the same as $r_{MS}$. The result with the primitive support functions is also shown for comparison. First, we focus on the results of the multi-site support functions whose coefficients are determined with the LFD method. The curve of multi-site support functions approaches to that of the primitive support functions as $r_{MS}$ increases. The energy difference from the primitive-support-function results are about or less than 0.2, 1.2 and 5.7 mhartree/atom for multi-site support functions ($r_{LD}$-$r_{MS}$) = (17.0-17.0), (8.0-8.0) and (5.0-5.0), respectively. We should note that the curvature of (5.0-5.0) is different from the others significantly. It indicates that the geometrical change of the target system affects the accuracy of the multi-site support functions largely when $r_{MS}$ is not large enough. Now, we consider the results of multi-site support functions whose coefficients are optimized after the LFD calculations. The optimization makes the multi-site-support-function curves closer to the primitive-support-function curve. The energy difference from the primitive-support-function results are about or less than 0.03, 0.8 and 1.7 mhartree/atom for multi-site support functions $(17.0\text{-}17.0)_{opt}$, $(8.0\text{-}8.0)_{opt}$ and $(5.0\text{-}5.0)_{opt}$. The coefficient optimizations improve the results of (5.0-5.0) significantly, not only for the energy difference but also for the curvature.

The bulk modulus $B_0$ and lattice constants $a_0$ in Table 1 are obtained by fitting the curves in Fig. 1 with Birch−Murnaghan equation. Table 1 also indicates that the multi-site support functions with large $r_{MS}$ provide the results closer to the primitive support functions. The differences between (17.0-17.0) and (8.0-8.0) are not large, which means that the multi-site support functions are converged in acceptable accuracy with $r_{MS} = 8.0$ bohr containing up to second neighbour atoms. (5.0-5.0) provides the large deviations from the primitive-support-function results, about 10 % for $B_0$ and 1 % for $a_0$. Although being smaller than the typical error from the use of LDA, 10 % error of $B_0$ may not be acceptable. However, these large errors by (5.0-5.0) are reduced dramatically by the optimization of the coefficients: the deviation of $B_0$ and $a_0$ by $(5.0\text{-}5.0)_{opt}$ is 5 % and 0.2 %, which are acceptable in most cases. Thus, if the PAO coefficients are optimized, the multi-site support function can provide reasonable accuracy with small cutoff ($r_{MS} = 5.0$ bohr) which includes only up to the nearest neighbour atoms. The use of small $r_{MS}$ will be important especially when we perform O($N$) calculations.

*3.2. Total energies and forces of a benzene molecule*

Next, we have performed calculations on a benzene molecule to investigate the accuracy of atomic force calculations with the multi-site support functions with and without the coefficient optimization.

Valence double $\zeta$ plus polarization (DZP) PAOs[20] has been used in the calculations. The numbers of primitive and multi-site support functions are 13 (=2s, 2p, d) and four for each C atom and five (=2s, p) and one for each H atom, respectively. The generalized gradient approximation (GGA)[21] functional is used and only $\Gamma$ point is taken into account in the diagonalization calculations. A benzene molecule is put in a supercell with the axes $(a, b, c) =$ (30, 30, 30) bohr.

Table 2 lists the total energies and forces of a distorted benzene molecule have been calculated with several multi-site support functions with several $r_{MS}$ and $r_{LD}$. The deviations from the results with the primitive support functions are listed. The geometry of the distorted benzene in $C_{2v}$ symmetry is made from a benzene in $D_{6h}$ symmetry with $r_{CC}$ = 2.646 bohr (= 1.4 Å) and $r_{CH}$ = 2.079 bohr (= 1.1 Å) by shifting a CH bond away from the center of the benzene ring by 1.0 bohr. The force on the carbon atom which is located opposite to the shifted carbon atom is shown in the table. (8.0-8.0) provides the energies and forces very close to those of the primitive support functions even without the coefficient optimization. On the other hand, (8.0-5.0) and (8.0-3.0) provide large deviations, about 0.002 and 0.027 hartree in energy and 0.01 and 0.08 hartree/bohr in force, respectively. These large deviations have been reduced by the coefficient optimization significantly. The deviations of (8.0-3.0)$_{opt}$ is 0.001 hartree/bohr, which is acceptable for most of the geometry optimizations and molecular dynamics simulations.

To check this, we have performed geometry optimization of the benzene molecule in $D_{6h}$ symmetry. The deviations of the optimized CC and CH bond lengths from the primitive results are summarized in Table 3. The deviations with both (8.0-8.0) and (8.0-5.0) are small even without the coefficient optimization, about or less than 0.01 bohr. The deviation of $r_{CC}$ with (8.0-3.0) is 0.04 bohr. This large deviation has been reduced to 0.005 bohr by the coefficient optimization. Thus, the coefficient optimization makes the geometry optimizations with small $r_{MS}$ be in reasonable accuracy. In the end of this section, we would like to make some comments. We have confirmed that the calculations of (5.0-5.0) and (5.0-5.0)$_{opt}$ give similar results with (8.0-5.0) and (8.0-5.0)$_{opt}$, respectively. However, we have failed to make PAO coefficients converged if we start the optimization from the result of (3.0-3.0). It is probably due to the poor accuracy of the initial PAO coefficients given by LFD with small $r_{LD}$, and it was necessary to set large $r_{LD}$ for the stable and accurate optimization with our present optimizer.

*3.3. Hydrated DNA system: Density of states*

We have also investigated the reproducibility of the density of states for a complex system. The target system in this section is B-DNA decamer 5'-d(CCATTAATGG)2-3' which contains 634 DNA atoms with 932 hydrating water molecules and 9 Mg counter ions, totalling 3439 atoms.[22] DZP PAOs[23] and GGA functional have been used and only Γ point has been taken into account in the calculations. 27883 primitive support functions are contracted into 7447 multi-site support functions. The significant reduction of the computational time with the multi-site support functions for this system has been reported in the previous study.[13] We have investigated the density of states (DOS) around Fermi levels obtained with the multi-site support functions (16.0-16.0), (11.0-11.0) and (8.0-8.0) with and without the coefficient optimization, which are presented in Figure 2. The DOSs are obtained by using a Gaussian broadening with the half-width of 0.003 hartree. The DOS obtained with the primitive support functions is also presented for comparison.

It is clearly shown that the use of large $r_{MS}$ and $r_{LD}$ improve the accuracy of DOS not only for the occupied states but also for the unoccupied states. (8.0-8.0) provides the deviation from the results of primitive support functions larger than (16.0-16.0) and (11.0-11.0). The deviations in the occupied region are reduced by the coefficient optimization, while those in the unoccupied regions are increased. This may because the coefficient optimization focuses only on the occupied states by minimizing the electronic energy. The deviation of (8.0-8.0)$_{opt}$ is comparable to that of (16.0-16.0), which mean that we can reach the high accuracy by using small cutoffs with the optimization instead of using large cutoffs.

*3.4. Hydrated DNA system: Energy convergence*

Another benefit of the optimization after LFD calculations with small cutoffs is to guarantee the variationality of energy minimizations. Figure 3 summarizes the changes of the total energy with respect to the update of the coefficients. These coefficient updates consist of the two-step update with the LFD method, which is explained in section 2, and subsequent numerical optimizations. The update with the LFD method has been continued until the total energy or the charge density converges with the threshold (e.g., $10^{-7}$ hartree in the calculations in Fig. 3). When $r_{MS}$ is large as (16.0-16.0) and (11.0-11.0), the update with the LFD method minimizes the total energies smoothly and the subsequent optimizations (16.0-16.0)$_{opt}$ and (11.0-11.0)$_{opt}$ reach convergence in a few steps. Furthermore, the use of large $r_{MS}$ reduces the number of the update steps to reach the threshold. On the other hand, (8.0-8.0) shows the difficulty of convergence. As shown in Fig. 3(b), the energy with (8.0-8.0) keeps fluctuating and does not converge to the threshold. This fluctuation can be removed if we stop the LFD update soon after the energy rise is found and start numerical optimization, shown as $(8.0-8.0)_{opt}$. Although the speed of convergence is slower than $(16.0-16.0)_{opt}$ and $(11.0-11.0)_{opt.}$, $(8.0-8.0)_{opt}$ also succeeded in reaching the convergence. Thus, the coefficient optimizations guarantee the variational energy minimization with multi-site support functions.

4. **Conclusions**

We have assessed the accuracy of the multi-site support functions[13] which have been recently introduced into CONQUEST in order to reduce the number of support functions while maintaining as much accuracy as possible. The multi-site support functions are constructed by taking linear combinations of the PAOs on both the target atom and its neighbour atoms within a cutoff. To determine the linear-combination coefficients, we have used the local filter diagonalization (LFD) method[14,15] and the optimization of the coefficients to minimize the total energy. The multi-site support functions with large cutoffs provide energies and forces with comparable accuracy with the original PAO results. When the coefficients are determined only with the LFD method, the multi-site support functions with small cutoffs provide large deviations from the original results. The optimization of the coefficients improves the accuracy of the multi-site support functions with small cutoff significantly. It has been also found that the convergence of the calculations with small cutoffs is also improved by the optimization. Thus, the coefficient optimization enables us to use small cutoff for multi-site support functions keeping reasonable accuracy, which will reduce the computational cost significantly.

**Acknowledgements**

AN is funded by ICYS-NAMIKI, NIMS. This work is supported by KAKENHI projects by MEXT (No. 22104005) and by JSPS (No. 25810015 and 26610120), Japan. This work is also supported by the Strategic Programs for Innovative Research (SPIRE) and the Computational Materials Science Initiative (CMSI), Japan. Calculations where performed in the Numerical Materials Simulator at NIMS, Tsukuba, Japan and the supercomputer HA8000 system at Kyushu-university, Fukuoka, Japan.

**AUTHOR INFORMATION**

**Corresponding Authors**

[*] Email: NAKATA.Ayako@nims.go.jp

[**] Email: MIYAZAKI.Tsuyoshi@nims.go.jp

**FIGURES.** (All figures are in actual size.)

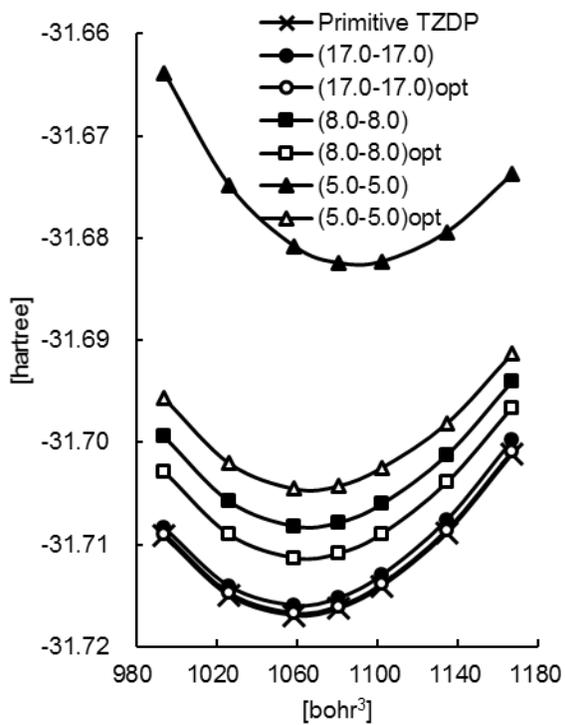

**Fig. 1** Energy-volume curves of crystalline silicon by multi-site support functions ($r_{LD}$-$r_{MS}$).

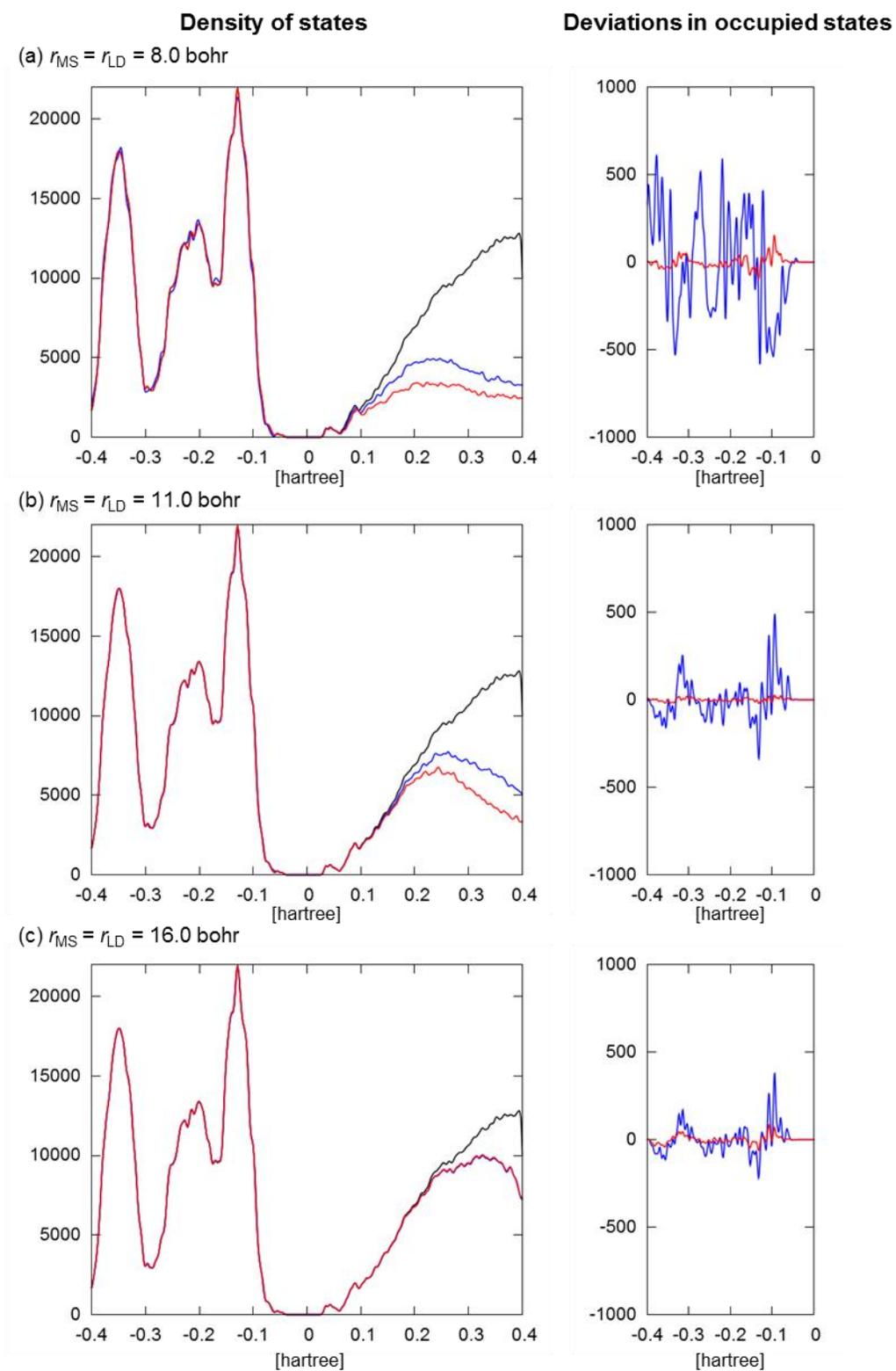

**Fig. 2** Density of states of the hydrated DNA system by primitive and multi-site support functions. Black, blue and red curves correspond to the results by primitive, multi-site (LFD) and multi-site (LFD + opt) support functions, respectively. $r_{LD}$ is set to be equal to $r_{MS}$.

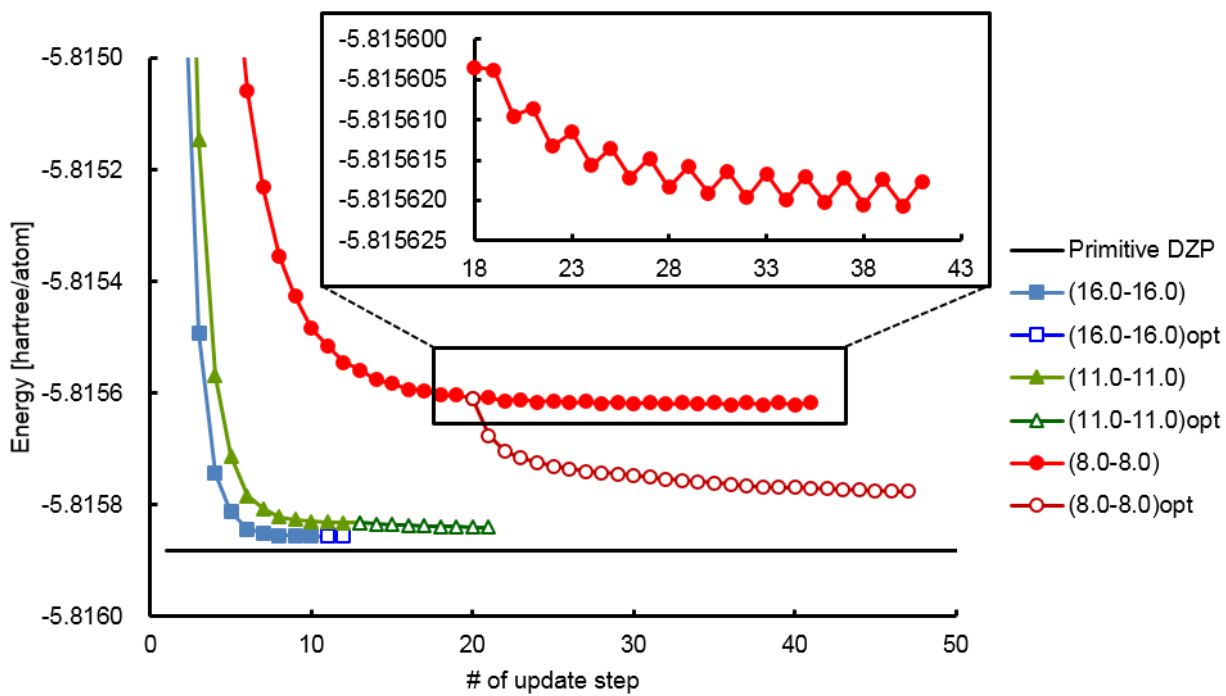

**Fig. 3** Energy convergence with respect to the update of the coefficients in multi-site support functions. Squares, triangles and circles correspond to the results of multi-site support functions ($r_{LD}-r_{MS}$) = (16.0-16.0), (11.0-11.0) and (8.0-8.0) respectively.

**TABLES.**

**Table 1** Bulk modulus $B_0$ [GPa] and lattice constants $a_0$ [bohr] of crystalline Si. The percent deviation (%Δ) from the results by primitive TZDP are also shown.

|  | $B_0$ | | %Δ of $B_0$ | |
|---|---|---|---|---|
|  | LFD | LFD + opt | LFD | LFD + opt |
| (5.0-5.0) | 110.1 | 94.9 | 9.8 | -5.4 |
| (8.0-8.0) | 98.7 | 99.5 | -1.6 | -0.8 |
| (17.0-17.0) | 100.9 | 100.3 | 0.6 | 0.0 |
| Primitive TZDP | 100.3 | | | |
|  | $a_0$ | | %Δ of $a_0$ | |
|  | LFD | LFD + opt | LFD | LFD + opt |
| (5.0-5.0) | 10.293 | 10.215 | 1.0 | 0.2 |
| (8.0-8.0) | 10.210 | 10.205 | 0.2 | 0.1 |
| (17.0-17.0) | 10.192 | 10.195 | 0.0 | 0.0 |
| Primitive TZDP | 10.195 (5.395 Å) | | | |

**Table 2** Differences of the total energies [hartree] and forces [hartree/bohr] of a distorted benzene molecule by multi-site support functions ($r_{LD}$-$r_{MS}$) from those by primitive support functions. Forces of the carbon atom opposite to the shifted carbon atom are listed.

|  | Energy difference | | Force difference | |
|---|---|---|---|---|
|  | LFD | LFD + opt | LFD | LFD + opt |
| (8.0-3.0) | 0.02652 | 0.00498 | 0.0833 | 0.0011 |
| (8.0-5.0) | 0.00194 | 0.00037 | 0.0101 | 0.0000 |
| (8.0-8.0) | 0.00007 | 0.00004 | -0.0003 | 0.0001 |
| Primitive DZP | -37.482270 | | -0.012835 | |

**Table 3** Differences of the bond lengths [bohr] of the optimized benzene molecule ($D_{6h}$) by multi-site support functions ($r_{LD}$-$r_{MS}$) from those by primitive support functions.

|  | rCC | | rCH | |
|---|---|---|---|---|
|  | LFD | LFD + opt | LFD | LFD + opt |
| (8.0-3.0) | -0.040 | 0.005 | -0.007 | -0.001 |
| (8.0-5.0) | 0.001 | 0.001 | -0.002 | 0.000 |
| (8.0-8.0) | 0.001 | 0.001 | 0.010 | 0.000 |
| Primitive DZP | 2.674 (1.415 Å) | | 2.098 (1.110 Å) | |

**For Table of Contents use only**

Text:

Newly introduced numerical optimization of multi-site support functions in the linear-scaling DFT code CONQUEST improves the accuracy and stability of the support functions with small cutoffs.

Graphical abstract (in actual size):

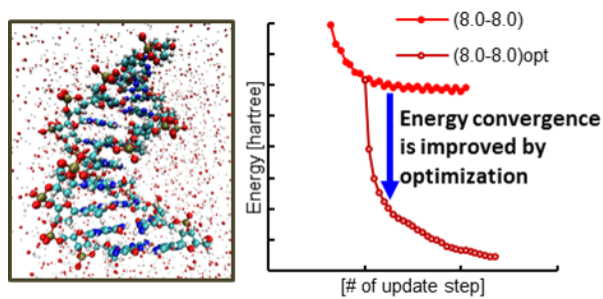

**Title of the paper**:

Optimized multi-site local orbitals in the large-scale DFT program CONQUEST

**Authors**: Ayako Nakata, David R. Bowler, Tsuyoshi Miyazaki